\documentclass[singlespacing]{elsart}
\usepackage{graphicx, amssymb}
\begin{document}

\begin{frontmatter}
\title{Parallel TREE code for two-component ultracold plasma analysis}
\author[lanl,ucdavis]{Byoungseon Jeon \corauthref{cor}}
\author[lanl]{Joel D. Kress}
\author[lanl]{Lee A. Collins}
\author[ucdavis]{Niels Gr{\o}nbech-Jensen}
\corauth[cor]{Corresponding author}
\address[lanl]{Theoretical Division, Los Alamos National Laboratory, 
Los Alamos, NM 87545}
\address[ucdavis]{Department of Applied Science, University of California, 
Davis, CA 95616}

\date{\today}

\begin{abstract} 
The TREE method has been widely used for long-range interaction {\it N}-body 
problems. We have developed a parallel TREE code for two-component 
classical plasmas with open boundary conditions and highly non-uniform charge
distributions. The program efficiently handles millions of particles
evolved over long relaxation times requiring millions of time steps.
Appropriate domain decomposition and dynamic data management
were employed, and large-scale parallel processing was
achieved using an intermediate level of granularity of domain decomposition
and ghost TREE communication. Even though the computational load is not fully
distributed in fine grains, high parallel efficiency was achieved for 
ultracold plasma systems of charged particles.
As an application, we performed simulations of an ultracold neutral
plasma with a half million particles and a half million time steps. 
For the long temporal trajectories of relaxation between heavy ions  
and light electrons, large configurations of ultracold plasmas
can now be investigated, which was not possible in past studies.
\end{abstract}


\end{frontmatter}

\newenvironment{mylisting}
{\begin{list}{}{\setlength{\leftmargin}{1em}}\item\scriptsize\bfseries}
{\end{list}}

\newenvironment{mytinylisting}
{\begin{list}{}{\setlength{\leftmargin}{1em}}\item\tiny\bfseries}
{\end{list}}

\newpage
\section{Introduction}
For {\it N}-body problems in gravitational and electrostatic phenomena,
significant computing resources are required due to the long range 
interactions. Despite advances in computing speed, it is still difficult to 
obtain sustainable results of realistically large configurations for many 
physical applications.

An ultracold plasma (UCP) is extensively studied in plasma physics, 
and is a typical example of long-range interactions for an open
boundary. Experimentally, it is generated by photoionization of laser-cooled 
heavy atoms, and the system has a low temperature ($\rm T =\mu K ~to ~ mK$)
compared to a conventional hot plasma ($\rm T = 10^3  ~to ~ 10^7 K$)
\cite{99killian}. By the initial disorder of particles, ions are heated, and 
active momentum transfer occurs between electrons and ions. We shall
study the behavior of charged particles and the physical properties of 
ultracold neutral plasmas.

Studying electron-ion coupling relaxation, it was found that 
several millions of time steps are required for ultracold
plasma evolution. This is a huge computing load, and the required time frame
restricts the size of problems that can be considered. We first implemented 
a molecular dynamics (MD) method with all pair-wise calculations of 
long-range interactions. However, such an approach, while accurate, scales
with the square of the number ($N$) of charged particles. Thus, it allowed
only simulations of $\rm 10^3 ~to~ 10^4$ particles over sufficiently long time
\cite{mazevet02,02kuzmin_prl,02kuzmin_plph}. In experiments,
common sizes of ultracold plasmas are reported as more than $10^6$ particles
\cite{99killian,simien04,chen04}, and larger computing capacity is 
therefore imperative for realistic simulations.

Consequently, an approximate method for evaluating the electrostatic forces is
necessary  in order to accelerate computing speed and increase the simulation
capacity. One of the candidates is the TREE method \cite{barnes86},
which has been widely used in astrophysical problems due to its $N\log N$ 
scaling of the computing cost. The basic idea is that particle interactions 
are calculated explicitly at close range while effective, averaged properties 
are considered for far-field interactions. Gravitational problems have a 
potential of  $1 / r$, where $r$ is the distance between two points, and are 
similar to a Coulomb system. Therefore we will be able to apply all the 
methods, which have been developed for astrophysics, to electron and ion 
interactions.

In addition to the serial TREE method, a parallel version has been constructed 
in order to distribute the computing load and thereby accelerate computation.
A specific shape of particle ensembles is considered from experimental data 
\cite{simien04,chen04}: charged particles are distributed non-uniformly, but
the overall shape is roughly symmetric. Particles are located inside of
a certain spherical volume, and this confirms simple and balanced domain 
decompositions for parallel computing of ultracold neutral plasmas. 
With the help of dynamic memory management and 
effective TREE communication, a highly efficient parallel code has been built.
Finally, we demonstrate that the program works well for 
ultracold plasma analysis. The following  describes  how the TREE method is 
implemented  for a two-component plasma (TCP) simulation, and how it is 
parallelized. Basic applications are also presented.

\section{TCP analysis} 
For a fully ionized plasma, we describe the interactions as Coulomb forces
between well-defined charged particles. However, because the electron mass 
is small compared to the ion, electrons will move faster than ions. This 
results in much smaller numerical time steps and longer relaxation times 
when simulating electrons compared to that of an ionic one-component plasma 
(OCP).

Since the dominating potential is Coulombic, each charged particle will 
interact with all the other charged particles and the interaction between
particles cannot be truncated at a characteristic distance. Using a TREE 
method, this extensive calculation can be achieved efficiently, and the 
computing load can be balanced. The force on each charged particle is 
calculated from the interaction with the TREE. Velocity and position of 
particles are updated using velocity Verlet \cite{allen} time integration 
once the force field has been evaluated.

\subsection{Modified Coulomb potential} 
The Coulomb pair potential between two point-charges, separated by the distance
$r$, is given by
\begin{equation}
V_{ij} = {C \over r}.
\end{equation}
\begin{equation}
C = {q_i q_j e^2 \over 4\pi \epsilon_0},
\end{equation}
where $q_i$ and $q_j$ are the fractional charges of particles $i$ and $j$. 
$e$ is the  unit charge, and $\epsilon_0$ is the vacuum permittivity. 
For the same kind of
particles (electron-electron and ion-ion), the forces are repulsive
whereas attractive forces are present for electron-ion interactions. 
Consequently, the bare Coulomb potential will result in an attractive 
singularity for the electron-ion pair at close distance ($r\approx 0$). 
However, quantum diffraction between electrons and ions in physical systems
prevents such point wise collisions, thus requiring a modification to 
the bare  Coulomb potential. Several modified Coulomb interactions have 
been proposed \cite{kelbg,deutsch77,pschiwul01}; 
we implemented the Kelbg potential:
\begin{equation}
V_{ei} =  {C \over r} \Big[ 1 -
\exp \Big( - {r \over \lambda_{ei}} \Big) \Big],
\label{eqn:kelbg}
\end{equation}
where $\lambda_{ei}$ is the thermal de Broglie wavelength
\begin{equation}
\lambda_{ei} =  \sqrt{2\pi \hbar^2 \over \mu_{ei} k_B T}.
\end{equation}
At close distances ($r<\lambda_{ei}$), the exponential term is dominant,
 and equation (\ref{eqn:kelbg}) asymptotes as $ {C \over \lambda_{ei} }  
\Big[ 1 -  {r \over 2 \lambda_{ei}} \Big]$.
The exponential term is negligible in the far field ($r\gg \lambda_{ei}$) 
where the equation converges to   $ {C\over r} $.
The thermal de Broglie wavelength plays a key role for close encounters; 
however, the temperature is not a well-defined quantity for these types of
systems. A better criterion results by setting $\lambda_{ei}$ to
a fixed value $r_s$ based on an averaged closest-approach or the lowest
Rydberg state allowed in recombination. Considering Rydberg states,
$\rm 20-180\AA$ will be the range of the parameter, and 
we use $\lambda_{ei} = \rm 100 \AA$ \cite{mazevet02} for our 
simulations. The basic properties of the TCP remain independent of this
cut-off parameter; its effect is merely to prevent very close encounters
that require very short time steps to resolve the dynamics accurately. For the
electron-electron and ion-ion interactions, the bare Coulomb potential is
applied. 

\section{TREE construction} 
Before explaining the parallel TREE method, we describe how to build
a TREE data structure for given particle sets. First, we assume a large cube 
that encloses all  particles. Then we evenly divide  the cube into 
eight small boxes, and each particle is associated with a box.
Further subdividing each of the small boxes into eight even smaller
pieces, all of the particles are associated with the smaller boxes again.
This refining is repeated until every box contains no more than one particle.
This hierarchical procedure leads to the tree shaped data structure.

The finest state of a TREE is called a {\bf leaf}. Above leaves, 
there exist large intermediate branches, referred to as {\bf twig}s 
\cite{tree}. A pseudo-particle is an effectively-charged particle of a twig, 
and the charge corresponds to the sum of the charges of subordinate twigs 
and leaves. The position of the pseudo-particle is interpolated from the lower 
twig or leaf positions depending on the charge ratio. 
When calculating the interactions between a particle and the TREE structure,
the distance between the particle and the twig, and the size of the box 
enclosing the corresponding twig will be inspected. If the relation satisfies 
certain criteria, the effective charge of the twig will be employed for 
the Coulomb potential. If not, the twig is opened until the criterion is
met. The opening criterion, due to Barnes as described in \cite{dubinski96}, 
reads
\begin{equation}
{s \over \theta } + \delta < d,
\end{equation}
where $s$ is the size of box, $d$ is the distance between a particle and 
a pseudo-particle, and $\delta$ is the distance between a pseudo-particle and 
the center of the box.

An opening criterion of $\theta < 1.0$ is commonly practiced.
A high opening criterion approximates force fields with upper branches of
the TREE - thus it accelerates simulations at the expense of accuracy.
In contrast, a low opening criterion demands fine calculations and 
results in higher accuracy with more computing resources. Depending on the
given problems and the required accuracy, an appropriate  $\theta $ will be 
determined. As mentioned above, neighboring particles will interact with 
each other but the interactions of remote particles will be replaced by
approximate pseudo-particles. Finally, the algorithm results in $N\log N$ 
scaling of computing cost, instead of $N^2$ scaling  of all pair-wise 
calculations. More detailed explanations about TREE construction can be found 
elsewhere \cite{barnes86,tree}.

\section{Parallelization} 
Even though the TREE method is computationally more efficient than direct 
evaluations of all pair-wise interactions for large configurations, 
a single processor calculation still 
remains inadequate for the task at hand. Therefore, we parallelized the 
TREE method onto multiple processors in order to distribute the computational 
load and accelerate computing speed. There have been several parallel schemes 
for TREE methods like hashed oct-tree \cite{warren93}, FLY \cite{becciani01}, 
and dynamic and adaptive domain decomposition \cite{miocchi02}.  
We have developed a simple domain decomposing parallel method using 
a ghost TREE communication. The problem domain is decomposed into coarse 
grains, rather than fine segments, that still allows for large scale parallel 
processing. Details are given below. The parallel routine was developed with 
the conventional message passing interface (MPI) library \cite{mpi}, 
allowing applications across most contemporary parallel computing platforms.

\subsection{Domain decomposition}
\begin{figure}
\centerline{\includegraphics[clip, scale=0.4, angle=0]{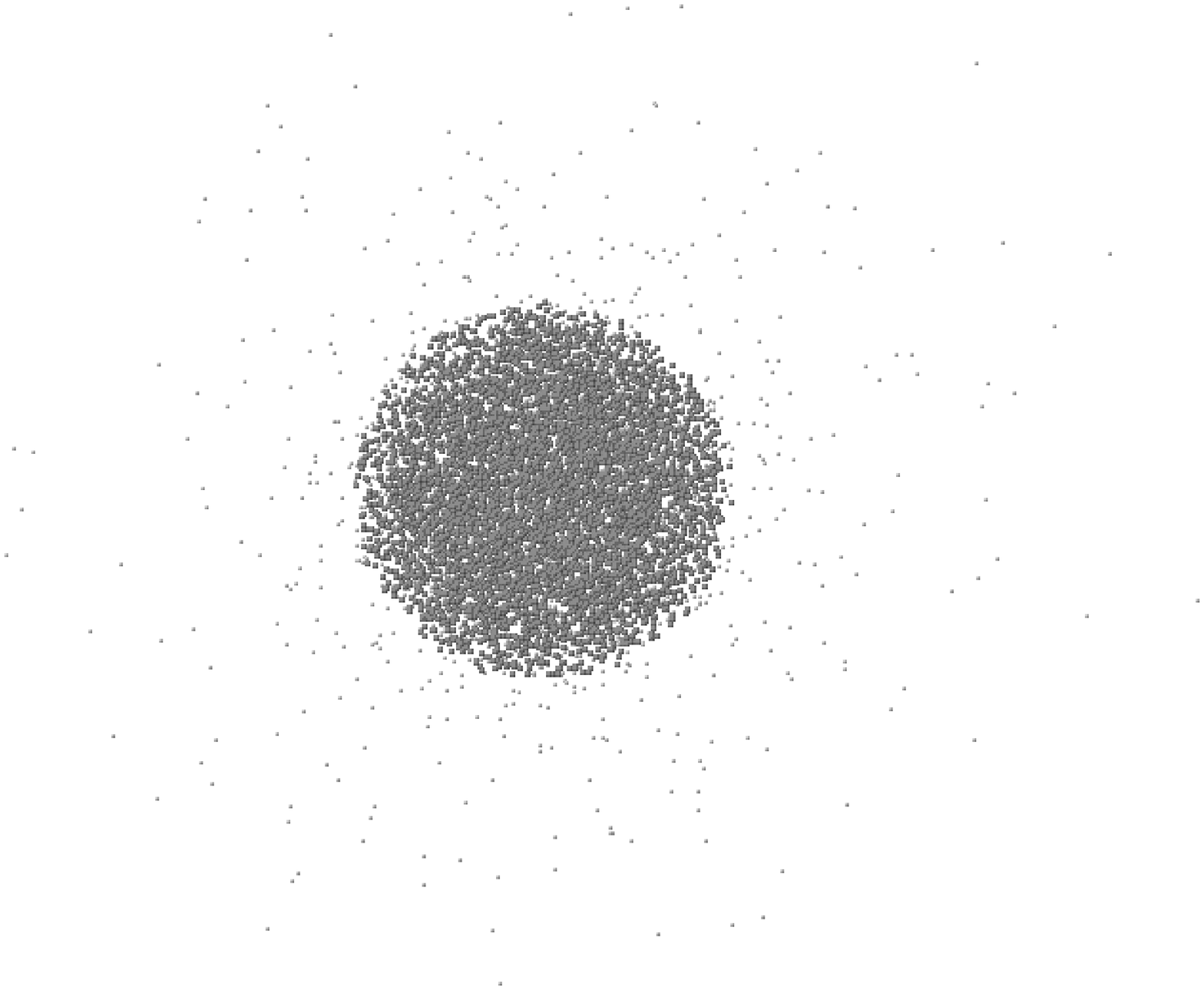}
\includegraphics[clip, scale=0.4, angle=0]{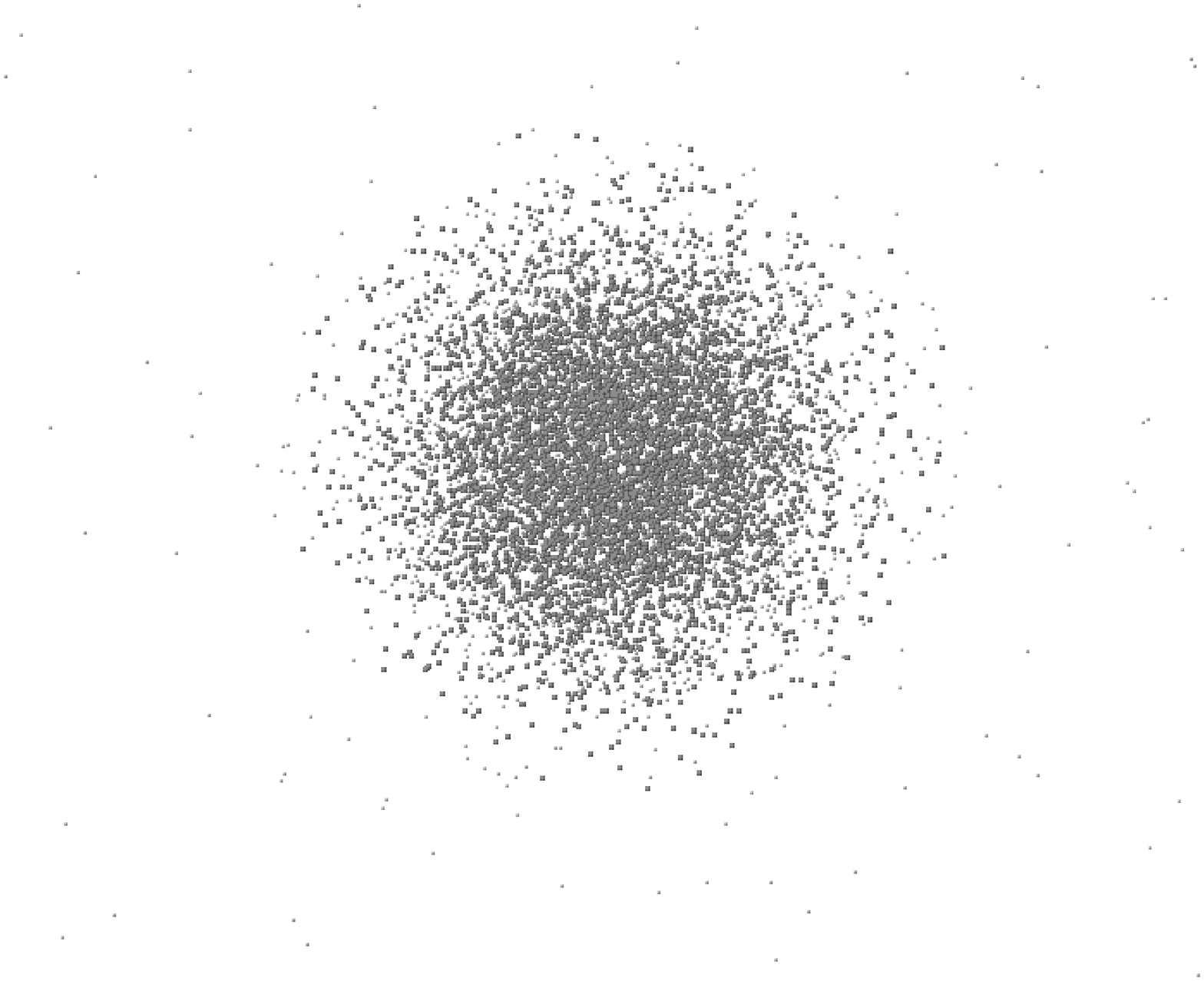}}
\caption{Electron($5\times 10^3$)/ion($5\times 10^3$) distribution - 0 ns (left) and 10ns (right)
 with reduced ion mass scheme.}
\label{fig:sph_shape}
\end{figure}

The basic configuration of the particles of ultracold plasmas maintains
over time a generally spherical and roughly symmetrical distribution having
though a highly non-uniform radial component 
as shown in Figure \ref{fig:sph_shape}. The center of the 
sphere can be placed at the origin, and we can assume that particles
are distributed randomly within the sphere. Using Cartesian
coordinates, an eight processor parallelization and domain decomposition can be
defined as shown in Figure \ref{fig:8cpu}. Each segment of geometrical space
will be designated to each processor, and they will communicate with 
each other. For more than eight processors, a finer domain decomposition
might be performed, but we still keep the eight segment domain decompositions.
We discuss the large-scale parallel computing below.

\subsection{Particle management} 
Due to the temporal evolution of particle positions during a simulation,
some particles will move between the spatial regions of the processors, and
each processor will have to manage the migrations.  For effective memory 
management, a dynamic memory allocation scheme has been implemented. 
After a Verlet time integration step, new positions of all the particles
are determined. Investigating the new positions, the particles which 
exceed the spatial range of their processor will be identified. Their 
information will be stored in a buffer, and each processor will remove 
these particles from memory. Then each processor sends and receives 
migrating particle data. For newly imported particles, their information is 
attached at the end of the memory block. 

\subsection{TREE communication} 
At the core of the parallel TREE algorithm is the communication between
processors, namely what and how to communicate. If we share the whole TREE
structure of all processors, communication will be operationally easy.
This scheme, though, requires huge data communication and management resources.
Hence, we need to be selective with communicating only necessary information
among processors.

Considering the basics of the TREE method, we need leaf information at close 
range but only twig (pseudo-particle) information for larger distances. 
This rule is also applied for parallel computing, and we employ 
pre-pruning before communication \cite{salmon_thesis,dubinski96}. 
If two processors are 
neighbors, a large fraction of the TREE will be necessary  whereas a 
small fraction will be enough if the processors are remote.

After copying the local TREE into a buffer, each processor begins to 
communicate with the other processors. Asynchronously, each 
processor sends its buffer and receives the TREE  information of other 
processors, called the ghost TREE \cite{wang06}. Using ghost TREEs, the 
interactions  with other particles of other processors will be evaluated.

With a given opening criterion, a certain amount of the local TREE will be 
sent to other processors, but we still need another step to convert the 
data. The TREE data are managed by FORTRAN pointers, but they are not
supported by a common MPI library to send/receive. Therefore, we need to 
convert them into combinations of MPI\_INTEGER and MPI\_REAL variables.
Not only the data component of each leaf and twig, 
but also the order and branch of the TREE should be communicated, and we have 
developed a simple data array to keep the order of the TREE structure. 

\begin{figure}
\centerline{\includegraphics[clip, scale=0.4, angle=0]{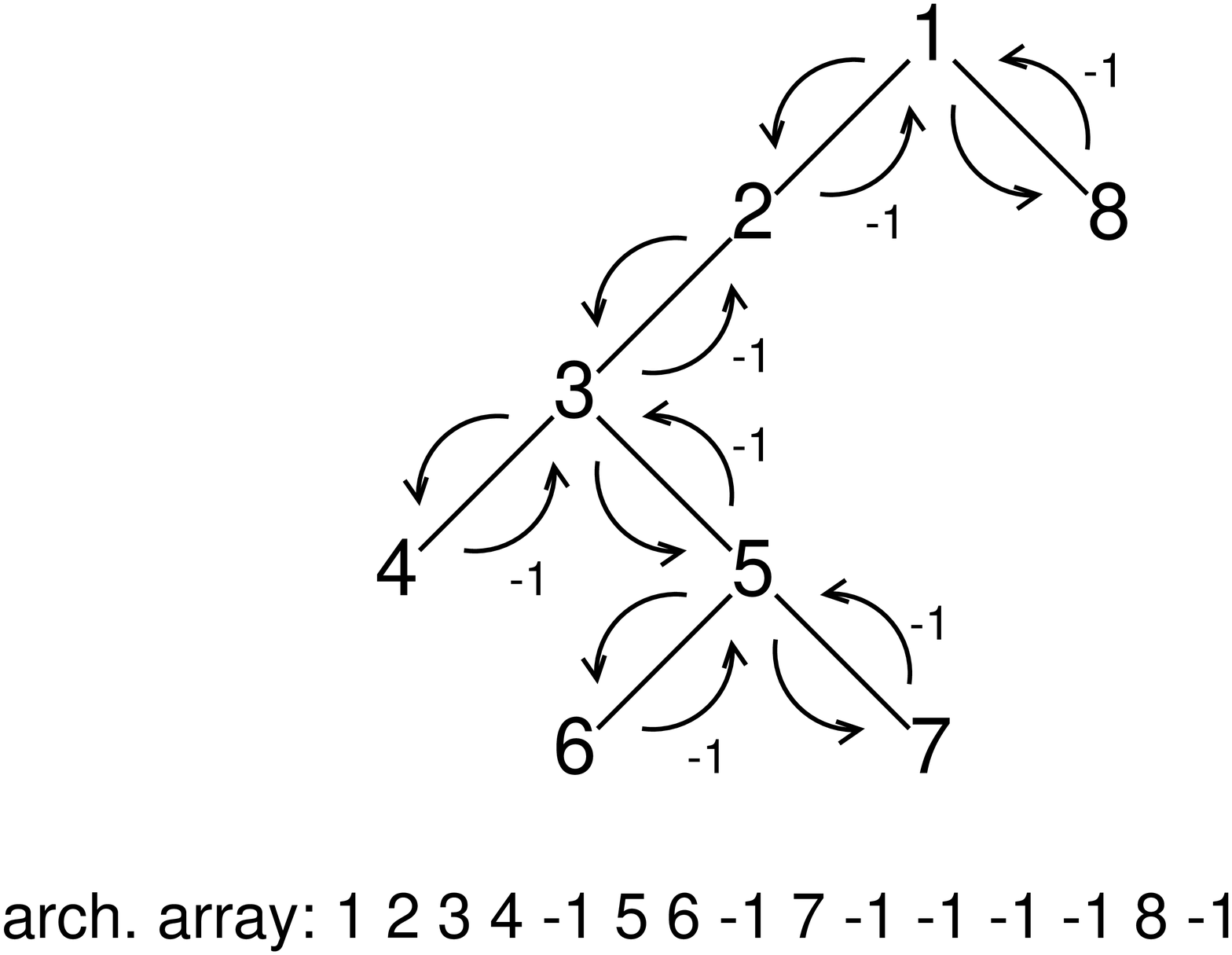}}
\caption{Architecture array for a sample TREE.}
\label{fig:arch}
\end{figure}

As shown in Figure \ref{fig:arch}, the TREE structure and order can be 
represented by an architecture series. Each natural number indicates the node 
array  of a pseudo-particle, which contains position and effective charge. 
A negative number (= -1) means a back step, and all of the branches and orders
of the TREE structure can be represented by the architecture series. 
Conversely, received ghost TREE data can be decoded into pointer forms 
using the received architecture series. Corresponding pseudo code is shown 
below:
\begin{mylisting}
\begin{verbatim}
RECURSIVE SUBROUTINE REBUILD(POINTER, N)
    N = N + 1
    ALLOCATE POINTER
    POINTER = NODE(ARCH(N))
    IF  ARCH(N+1)>0 THEN
       CALL  REBUILD(POINTER_Child_1, N)
       N = N + 1
       IF  ARCH(N+1)>0 THEN
           CALL  REBUILD(POINTER_Child_2, N)
           N = N + 1
       END IF
       ......
                     
       IF  ARCH(N+1)>0 THEN
           CALL  REBUILD(POINTER_Child_8, N)
           N = N + 1
       END IF
    END IF
END SUBROUTINE
\end{verbatim}
\end{mylisting}

\subsection{Large scale parallel processing} 
\begin{figure}
\centerline{\includegraphics[clip, scale=0.4, angle=0]{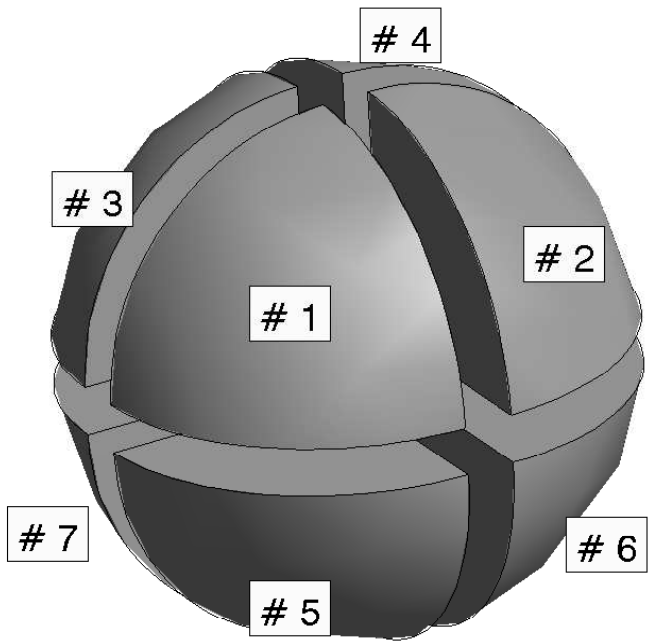}}
\caption{Schematic view of eight-segment domain decomposition.}
\label{fig:8cpu}
\end{figure}

As mentioned earlier, the simulations of ultracold plasma evolution demand
a large number of time steps, typically of the order of $10^5-10^6$. 
This time frame and the available computing power mediate the largest 
configuration which we can handle as around a million particles. The fine 
granularity of {\it N} domain decompositions of {\it N} processors will 
reduce the size of the local TREE to a system of less than $10^4$ particles 
for which the TREE efficiency becomes poor \cite{tree}, for large scale 
parallel computing. But the intermediate granular 
parallelism, which employs an intermediate granularity by  administrating 
local TREEs as a unit of a single segment of Figure \ref{fig:8cpu},  
still keeps the local TREE as $10^5$ particles or more than that, providing 
higher efficiency. Also we can expect easy book-keeping and simple data 
handling.

We maintain eight piece domain decompositions, as shown above, and apply
processors of  multiples of eight. Then each segment of Figure \ref{fig:8cpu}
will have the same number of local processors.
All particles
of each segment are distributed evenly for each local processor  for load
balance. When the local TREE of each segment is built, all local processors
of each segment swap and share the position of particles of the segment. 
Then the local TREE is built at every local processor, without parallelism.
However, particle interactions with the local TREE, which consume most of the
computing resources,  are calculated in parallel. 

For the communication between segments, one of the local processors will 
be assigned as the {\bf head processor}. After calculating particle 
interactions with the local TREE, the head processor of each segment will 
copy and send the local TREE  to other segments while  receiving ghost 
TREEs from the corresponding segments. This communication is performed 
only by the  head processor, and the communication burden can thereby be 
reduced. After communication between head processors, the received ghost 
TREEs are distributed again on the local processors inside the segment.
Ghost TREEs are handled as the same way of local TREEs. The decoding is 
done in serial but particle interactions are done in parallel.

\begin{figure}
\centerline{\includegraphics[clip, scale=0.25, angle=0]{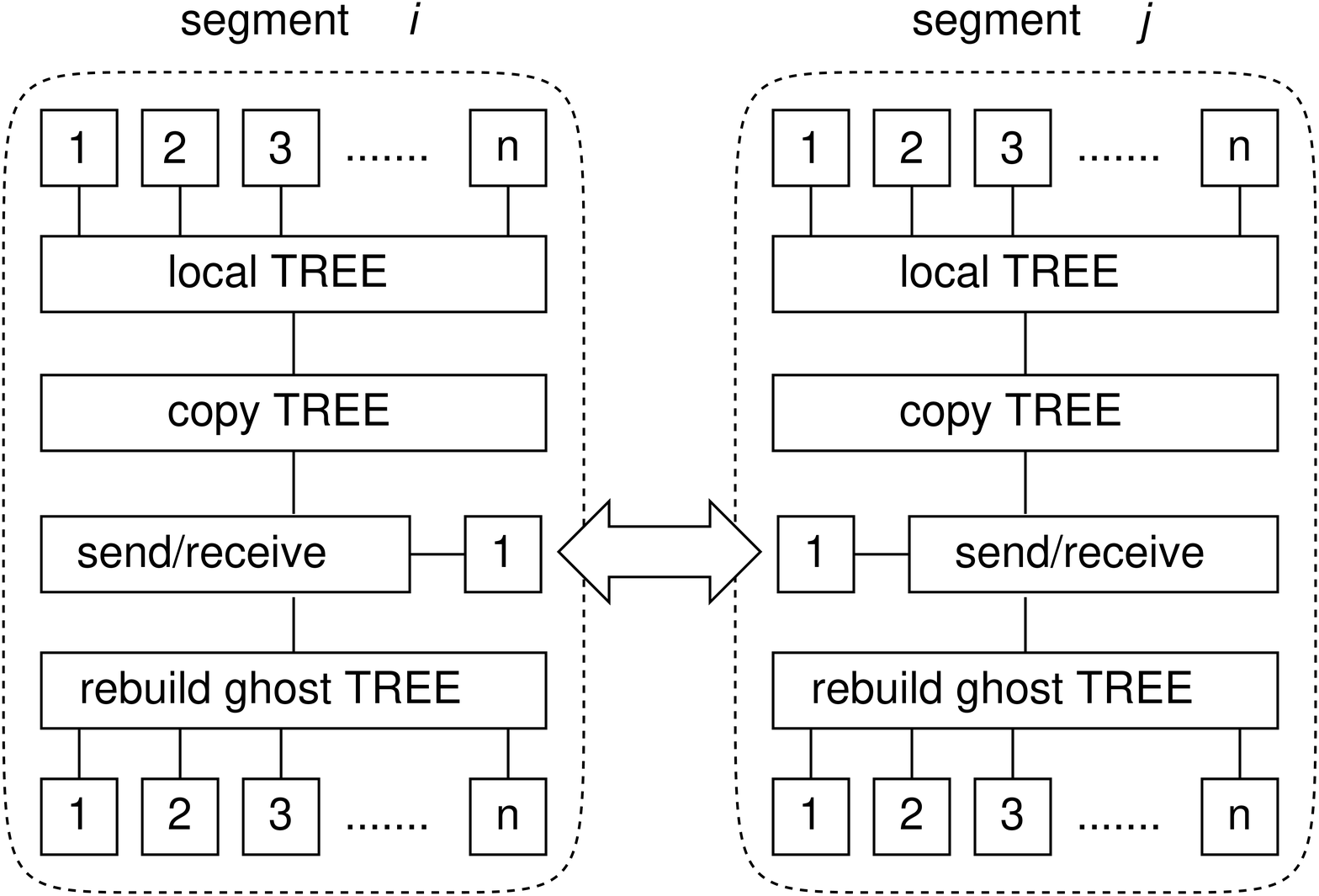}}
\caption{Calculation and communication flow of each segment.}
\label{fig:segment}
\end{figure}

The schematic flow of each segment is shown in Figure \ref{fig:segment}. 
Using this method, multiples of eight processors can participate in the 
parallel computing. Even though this method does not fully distribute 
the computing load with fine granularity, it allows for easy  book-keeping 
and reduces the communication load. Although the experimental charge 
distributions of ultracold plasmas are non-uniform, they retain an overall 
spherical and roughly symmetrical appearance.
Therefore eight-segment domain decomposition 
will divide the problem domain quite evenly, and we can expect good load 
balance from the beginning. Also divided particles are distributed uniformly 
for the local processors of each segment. During simulations, some of 
particles will move between segments as discussed above, and new particles 
are sent to a local processor, which has the least particles in the segment. 
These procedures maintain good load balance for each processor.

In addition to the message passing interface, OpenMP \cite{openmp} has been 
implemented in order to utilize shared memory processing. Basic operation 
of the code is executed in a single thread, but particle interactions with 
the TREE  are forked along multiple threads and computation is thereby 
accelerated. 

\section{Computational experiments} 
As an example to illustrate the developed method, we investigate the dynamics 
of electrons and ions with initial conditions of previous work  
\cite{mazevet02}. The configuration has same electron/ion density 
($\rho_i= 4.32\times 10^9 \rm /cc$), and particles  are distributed inside 
a sphere. The initial temperature is 3 K for electrons and  
$1\times 10^{-6}$ K for ions. A reduced ion mass
simulation \cite{mazevet02}, which artificially reduces the ion mass 
from 131 to 0.01 amu, is employed in order to expedite momentum transfer 
between electrons and ions. A time step of 20 fs and $5\times 10^5$ time steps 
are employed for the plasma evolutions.

\begin{figure}
\centerline{\includegraphics[clip, scale=0.45, angle=0]{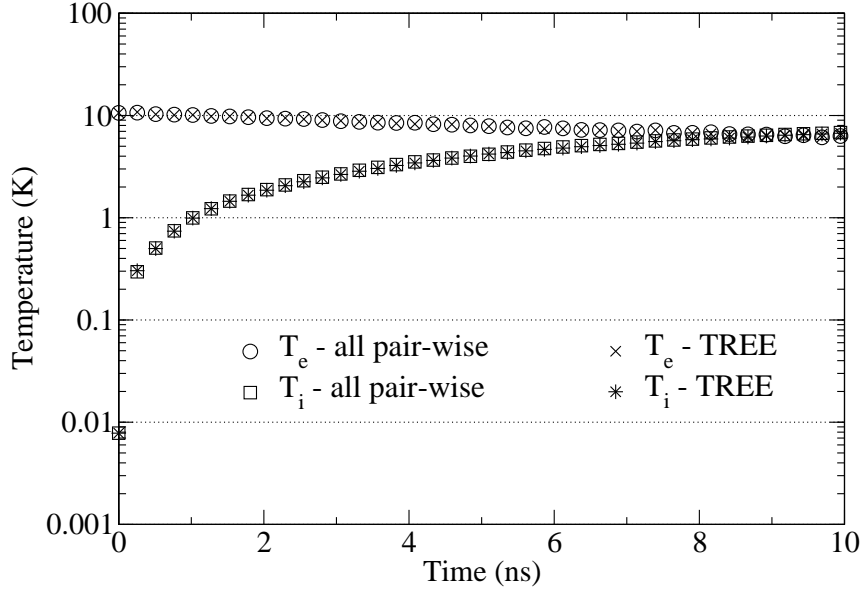}}
\caption{Electron ($\rm T_e$) and ion ($\rm T_i$) temperature by the all 
pair-wise evaluation and the TREE method ($5\times 10^3$ electrons and 
$5\times 10^3$ ions).}
\label{fig:compare}
\end{figure}

With these configurations, a UCP system of $5 \times 10^3$ electrons ($q=-1$) 
and  $5 \times 10^3$ ions ($q=+1$) has been tested 
with all pair-wise and TREE methods in order to evaluate the developed
parallel TREE code. The test machine is grendel, a linux cluster at Los 
Alamos National Laboratory, which is composed of 126 nodes of dual 2.4GHz 
Xeon processors with Myrinet interconnections. 2Gb of memory is equipped for
each node. Los Alamos MPI (LAMPI) library
and OpenMP API of intel fortran compiler 9.1.036 were employed to compile
the code.  As shown in Figure \ref{fig:compare}, the average kinetic energy 
($=3k_B T/2$) of the electrons ($\rm T_e$) and ions ($\rm T_i$) 
approach the same asymptote. Therefore, we conclude that the developed 
TREE code reproduces the energy transfer well. As for computing 
cost, the all pair-wise method took 62 hours with 16 processors while
the TREE code ($\theta=0.4$) used 26 hours with 64 processors. 
Even though the TREE code  efficiency is less than the all pair-wise 
calculations  for this small set of particles,
the TREE code scales as  $N\log N$, while the all pair-wise  method scales 
as  $N^2$ as discussed  above. Consequently, the TREE code is imperative
for larger systems involving  millions of particles. For a one-million 
particle set, the all pair-wise method requires $10^4$ times more 
computing resources than above, and this is not practical. 

To determine parallel performance, larger sets of particles were tested.
Also several opening criteria have been tested, and the corresponding errors
were found. Finally, simple applications of large ultracold plasmas are given
below.

\subsection{Parallel performance} 
Parallel performance  was tested with $10^5$ particles 
($5\times 10^4$ electrons and $5\times 10^4$ ions).
Two kinds of opening criteria were tested, 
$\theta = 0.4$ and $ 0.6$.  The results below were achieved on flash, 
a linux cluster at Los Alamos National Laboratory, which is composed of 
300 nodes per segment, with dual Opteron processors of 2.0-2.4GHz and 
Myrinet interconnections. 8-16Gb of memory is equipped for each node.
Los Alamos MPI (LAMPI) library and OpenMP API of intel fortran compiler 
9.1.037 were employed to compile the code.

\begin{figure}
\centerline{\includegraphics[clip, scale=0.45, angle=0]{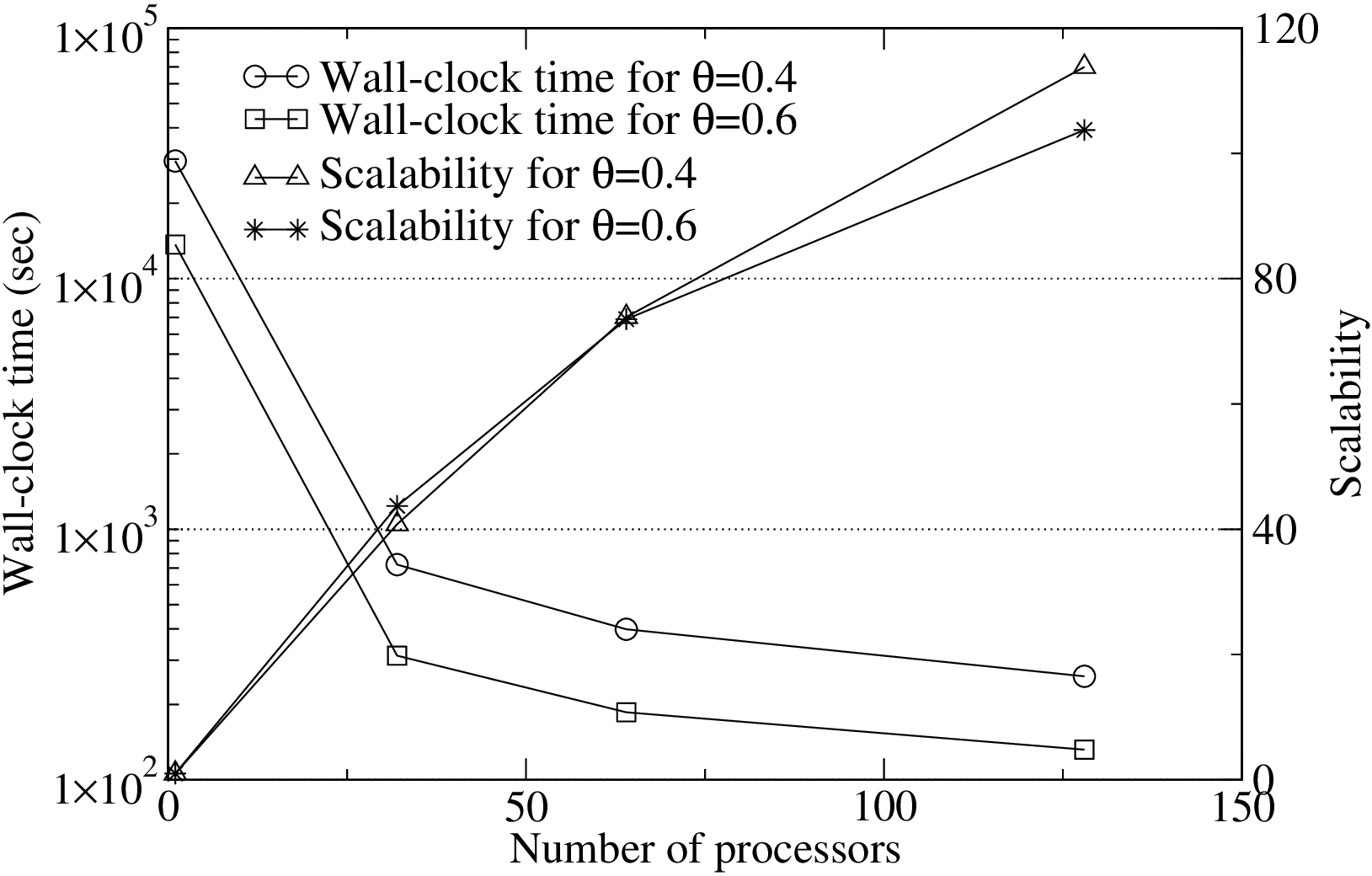}}
\caption{Performance using MPI and OpenMP hybrid parallel computing for 
$5\times 10^4$ electrons and $5\times 10^4$ ions. Based on the result of a 
single processor calculation, scalabilities were determined.}
\label{fig:parallel}
\end{figure}

The results are summarized in
Figure \ref{fig:parallel}. Scalability is the ratio between wall-clock time
of a parallel execution and the time of the corresponding serial code. 
With a dual processor machine, 16, 32, and 64
nodes were used while the number of processors are 
32, 64, and 128 respectively. For 128 processors, better than 80\% 
parallel efficiency was achieved for both opening criteria.

\begin{figure}
\centerline{\includegraphics[clip, scale=0.45, angle=0]{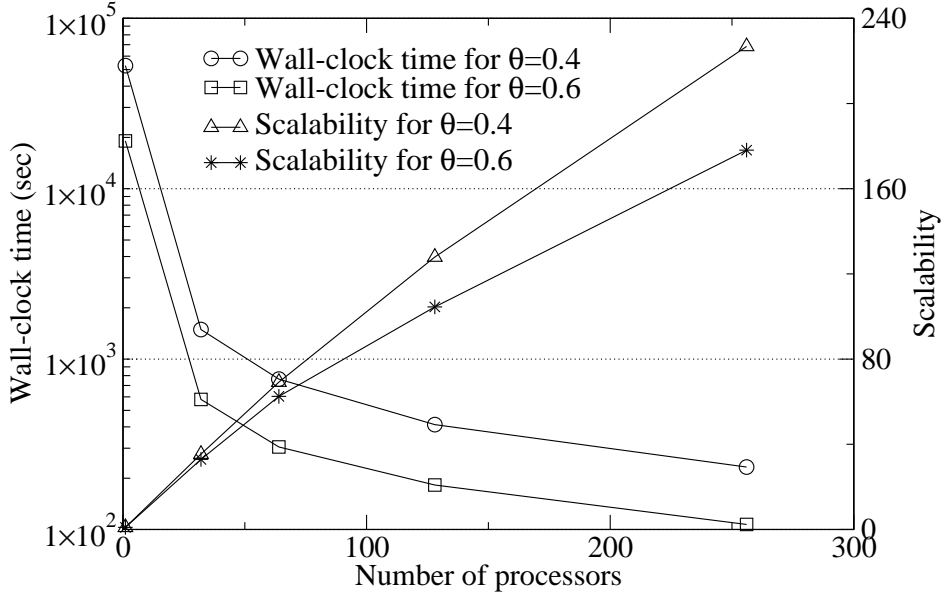}}
\caption{Performance using MPI and OpenMP hybrid parallel computing for 
$5\times 10^5$ electrons and $5\times 10^5$ ions.  Based on the result of 
a single processor calculation, scalabilities were determined.}
\label{fig:parallel2}
\end{figure}

Also a $10^6$ particle set ($5\times 10^5$ electrons and $5\times 10^5$ ions) 
was tested to study parallel efficiency. Because the size is 10 times
larger than above, we simulated with up to 256 processors; the  results
are summarized in Figure \ref{fig:parallel2}. Compared to a small system, the
effect of the opening criteria is clear. A low opening criterion imposes larger
computing loads, showing efficient parallel computing, and more wall clock 
time is consumed. A high opening criterion shows less parallel efficiency,  
but completes a calculation faster. With current computing resources and
$\theta = 0.6$, one million particle system with one million time steps
takes approximately 700 hours with 256 processors.

\begin{figure}
\centerline{\includegraphics[clip, scale=0.45, angle=0]{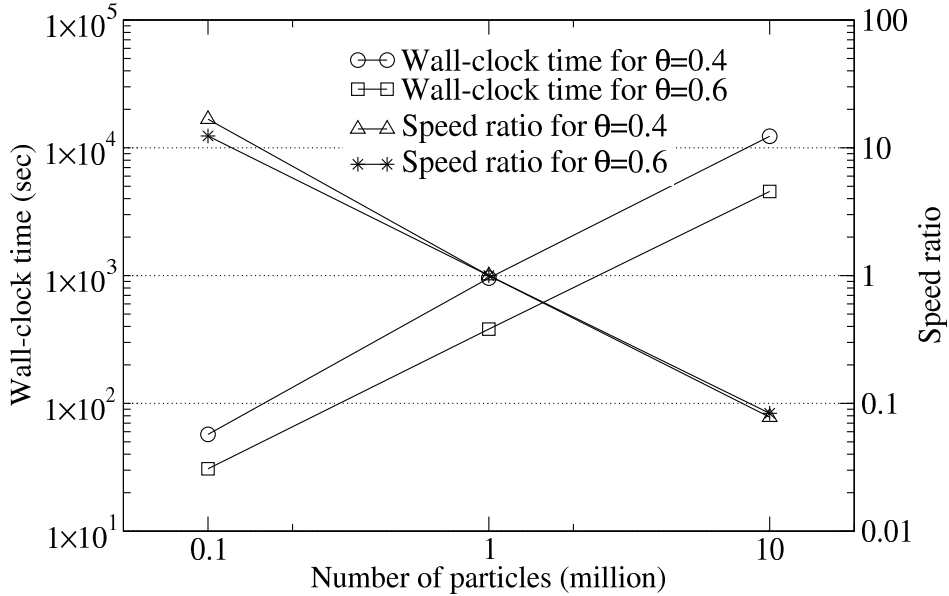}}
\caption{Performance curves of the parallel TREE code for different sizes.
Speed ratio is the wall-clock time ratio for the 1.0 million particle
results.}
\label{fig:size}
\end{figure}

The effect of configuration size was tested. Systems with 0.1, 1.0, and 
10 millions of particles (half electrons and half ions) 
have been tested with different number of nodes. 100 time
steps were tested and wall-clock times were examined per each section 
as shown in Table \ref{table:size}.
First, using 8 nodes of 16 processors, basic domain decomposition was tested.
Even though 2 processors of a single node share the same segment, parallel
computing is done by OpenMP only. Consequently, a single segment is taken 
care of by a single node only, and the result will serve as a reference for 
the bare eight segment domain decomposition, without the intermediate 
parallelism. Most of the computing resource is devoted to particle
interactions and updates. By pre-pruning, the load of communication and 
ghost TREE management is affected by the opening criterion.  The relative 
cost of each section is quite consistent for the size variations of systems. 
64 nodes of 128 processors were also applied, 
employing intermediate parallelism. Because 8 nodes share
a single segment, the cost of communication and TREE management is relatively 
high compared to 8 node calculations. But it decreases as the number of 
employed particles increases, providing better efficiency. 
For the 64 node results, wall-clock times and speed ratios, calculated 
relative to the 1 million particle results, are shown in Figure \ref{fig:size}.
For the change of problem sizes, speed ratios show $N\log N$ scaling.

\subsection{Opening criterion study} 
\begin{figure}
\centerline{\includegraphics[clip, scale=0.45, angle=0]{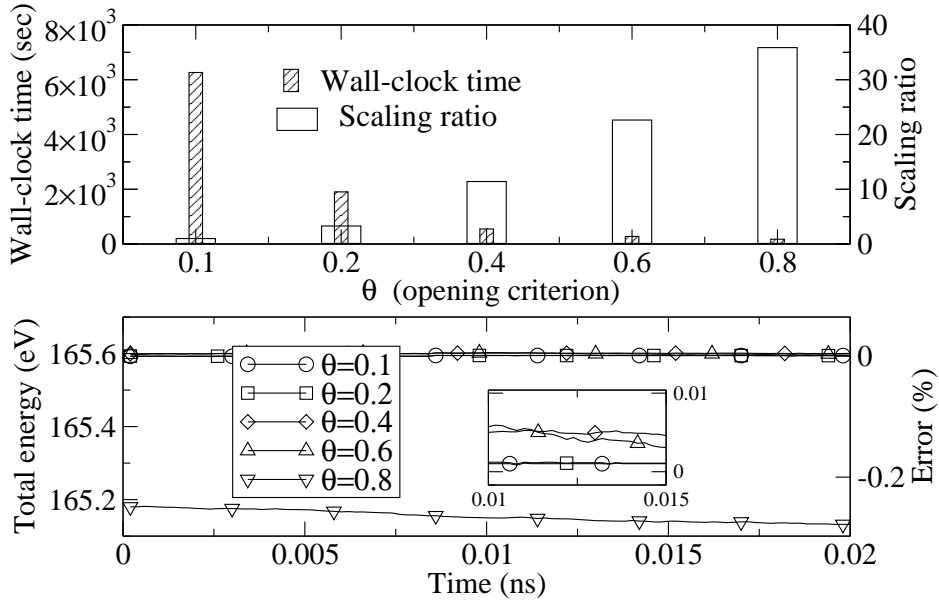}}
\caption{Performance and energy curves in terms of opening criteria. Scaling
ratios are the wall-clock time ratios with respect to $\theta=0.1$, whereas
the reference of error estimations was from all pair-wise calculations. 
The inset shows fine differences for $\theta=0.1-0.6$, 
where x-axis is time (ns) and y-axis is error(\%).}
\label{fig:sdratio}
\end{figure}

TREE code performance and accuracy depend on the opening criterion.
$5\times 10^4$ electrons and  $5\times 10^4$ ions were tested with 
several opening criteria in order to determine accuracy and efficiency 
differences under same computing conditions. 1,000 time steps were employed
with 64 nodes of 128 processors on flash cluster.
Figure \ref{fig:sdratio} provides the results where the error of total 
energy was measured against the all pair-wise results, and scaling ratio is 
relative wall-clock time for $\theta = 0.1$. For $\theta \le 0.6$, 
energy errors are quite small ($< 0.01 \%$) and converge to the result of 
all pair-wise calculations for smaller $\theta$, accompanying fluctuations.
However, computing cost increases quite quickly
as $\theta$ decreases; $\theta = 0.1$ demands three times the computing time 
of $\theta = 0.2$, and $\theta = 0.2$ costs three times that of $\theta = 0.4$.
Therefore a balance between accuracy and efficiency needs to be determined.
Also the energy error increases during the long temporal relaxation of charged
particles, and $\theta \le 0.7$ will be required for plasma relaxation 
simulations of our study.

\subsection{Evolution of two-component ultracold plasmas}
\begin{figure}
\centerline{\includegraphics[clip, scale=0.45, angle=0]{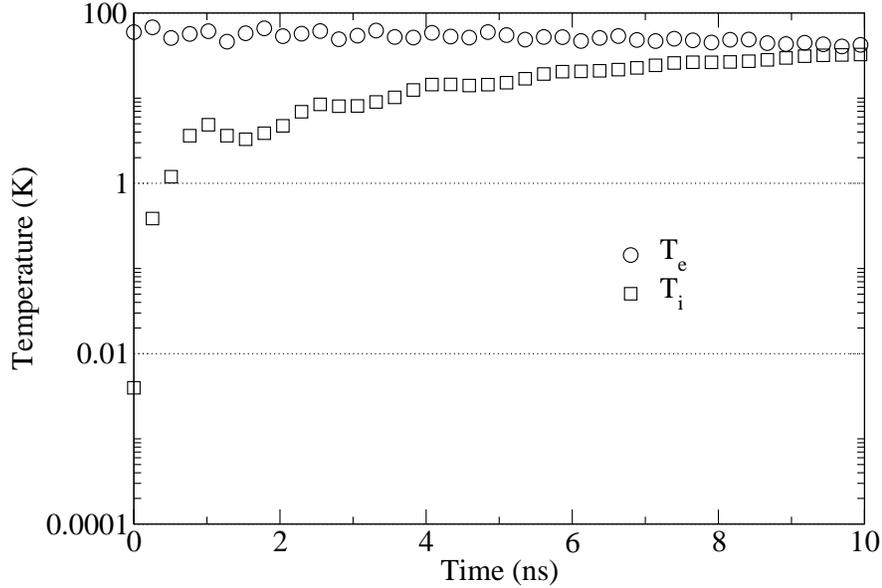}}
\caption{Evolution of electron and ion temperatures with $2.5\times 10^5$ 
electrons and $2.5\times 10^5$ ions over $5\times 10^5$ time steps.}
\label{fig:temp}
\end{figure}

Using the above mentioned initial conditions and reduced ion masses,
electron and ion temperatures were investigated up to 10 ns with 
$2.5\times 10^5$ electrons and $2.5\times 10^5$ ions, along  $5\times 10^5$
time steps. 
Figure \ref{fig:temp} provides the results, which are consistent 
with previous work \cite{mazevet02}. We can see that the average ion 
temperature increases rapidly for early times and saturates after 2 ns. 
Compared to the previous work (500 electrons and 500 ions), we employed much 
more particles, resulting in higher initial potential energy and more 
heating of the ions. We found periodic oscillations in the electron 
temperature evolution, which might be the effect of an electron plasma wave.
This will be studied extensively in future work \cite{jeon}. 

\section{Concluding remarks}
To analyze the relaxation behavior of ultracold plasmas, a TREE code
has been developed. Also, a parallel implementation has been conducted 
in order to allow for simulations of sufficiently large systems over
sufficiently long time. Dynamic data management was implemented for
efficient memory allocation of data communication, and eight-segment domain
decomposition was applied with a ghost TREE method. A parallelism of an 
intermediate granularity was developed for large scale parallel processing, 
and optimized for systems with millions of charged particles. 
Finally, the developed code has been found to
work well with the given ultracold plasma systems and efficient 
parallel performance has been demonstrated for various configurations. 

As discussed above, a simulation of two-component ultracold plasmas requires 
significant computing resources. Previously, only $\rm 10^3~to~10^4$ particles 
could be employed because of the long temporal trajectories of ultracold 
plasma relaxation. Here, with the developed  parallel TREE code, we have 
increased the computing capacity  up to  $\rm 10^6$ interacting particles. 
This is a crucial achievement in the study  of ultracold plasma dynamics.
Much larger and realistic configurations of ultracold plasmas can now
be investigated.

\section*{Acknowledgments}
We thank J. E. Barnes (University of Hawaii), M. Challacombe (LANL), 
and Z. Wang (LANL) for valuable discussions about the TREE method. 
Requests to the source code should be addressed to authors.
This work was supported by the Advanced Simulation Computing program of 
Los Alamos National Laboratory, and was carried out under the auspices 
of the National Nuclear Security 
Administration of the U.S. Department of Energy at Los Alamos National 
Laboratory under Contract No. DE-AC52-06NA25396.


\clearpage

\begin{table}[b]
\caption{Computing resource (in seconds) for parallel TREE simulations in 
terms of employed particles and opening criterion. {\it N} is the number
of particles. $\theta$ is the opening criterion. $n$ is the
number of computing nodes.}
\begin{center}
\renewcommand{\arraystretch}{1.0}
\begin{tabular}{r|r|r|r|r|r|r}
\hline
\hline
 & & & \multicolumn{4}{|c} {section} \\
\cline{4-7}
& & & particle & & &  \\
& & & interactions & local TREE & ghost TREE & commu- \\
{\it n}~ &{\it N~~~~} & $\theta$~  & and updates & management & management & nication\\

\hline
8 & 0.1 million & 0.4  & 314.0 & 1.5 & 17.8 & 2.5 \\
  &             & 0.6  & 139.2 & 1.5 &  8.0 & 1.7 \\
\cline{2-7}
  & 1 million   & 0.4  & $6.6\times 10^3$ & 20.4 & 346.9 & 9.9\\
  &             & 0.6  & $2.4\times 10^3$ & 20.4 & 141.0 & 6.3 \\
\cline{2-7}
  & 10 million  & 0.4  & $9.1\times 10^4$ & 255.3 & $2.1\times 10^3$ & 42.1 \\
  &             & 0.6  & $3.1\times 10^4$ & 254.6 & 909.2 & 26.9 \\
\hline
64 & 0.1 million & 0.4  & 40.1     & 1.9   & 7.5   & 7.2 \\
   &             & 0.6  & 18.2     & 2.5   & 3.4   & 4.7 \\ 
\cline{2-7}
   & 1 million   & 0.4  & 826.2    & 22.9  & 71.9  & 29.7  \\
   &             & 0.6  & 305.0    & 21.9  & 33.3  & 19.5 \\
\cline{2-7}
   & 10 million  & 0.4  & $1.1\times 10^4$ & 270.6 & 469.5 & 130.0  \\
   &             & 0.6  & $4.0\times 10^3$ & 272.3 & 205.7 & 85.1 \\
\hline
\hline
\end{tabular}
\end{center}
\label{table:size}
\end{table}

\clearpage

\end{document}